\documentclass[amssymb,twocolumn,aps]{revtex4}
\usepackage{graphics,color,graphicx,amsmath}
\usepackage{subcaption}
\usepackage{natbib}
\usepackage{verbatim}
\usepackage{graphicx}
\usepackage[above,below]{placeins}
\usepackage{float}
\usepackage{tabularx}
\usepackage{times}
\usepackage{blindtext}
\usepackage{url}
\usepackage{rotating}
\usepackage{mathtools}
\usepackage{threeparttable}

\begin{document}

\begin{titlepage}

  \title{Multi-institutional assessment of Peer Instruction implementation and impacts using the Framework for Interactive Learning in Lectures}

  \author{Ibukunoluwa Bukola$^{1}$, Meagan Sundstrom$^{1}$, Justin Gambrell$^{2}$, Olive Ross$^{3}$, Adrienne L. Traxler$^{4}$, and Eric Brewe$^{1}$}
    \affiliation{$^{1}$Department of Physics, Drexel University, Philadelphia, Pennsylvania 19104, USA\\
    $^{2}$Department of Computational Mathematics, Science and Engineering, Michigan State University, East Lansing, Michigan 48824, USA\\
    $^{3}$Laboratory of Atomic and Solid State Physics, Cornell University, Ithaca, New York 14853 USA\\
    $^{4}$Department of Science Education, University of Copenhagen, Copenhagen, Denmark}


  \begin{abstract}
    Substantial research indicates that active learning methods improve student learning more than traditional lecturing. Accordingly, current studies aim to characterize and evaluate different instructors' implementations of active learning methods. Peer Instruction is one of the most commonly used active learning methods in undergraduate physics instruction and typically involves the use of classroom response systems (e.g., clickers) where instructors pose conceptual questions that students answer individually and/or in collaboration with nearby peers. Several research studies have identified that different instructors vary in the ways they implement Peer Instruction (e.g., the time they give students to answer a question and the time they spend explaining the correct answer); however, these studies only take place at a single institution and do not relate the implementation of Peer Instruction to student learning. In this study, we analyze variation in both the implementation and impacts of Peer Instruction. We use classroom video observations and conceptual inventory data from seven introductory physics instructors across six U.S. institutions. We characterize implementation using the Framework for Interactive Learning in Lectures (FILL+), which classifies classroom activities as interactive (e.g., clicker questions), vicarious interactive (e.g., individual students asking a question), or non-interactive (e.g., instructor lecturing). 
    Our preliminary results suggest that instructors who use both interactive and vicarious interactive strategies may exhibit larger student learning gains than instructors who predominantly use only one of the two strategies.
    \clearpage
  \end{abstract}

  \maketitle
\end{titlepage}

\section{Introduction}

Extensive research demonstrates that active learning is better than traditional lecturing at improving students' conceptual physics knowledge~\cite{hake1998interactive}. Correspondingly, education researchers have deemed that ``there is no need to conduct additional research on this point''~\cite[p. 9]{dancy2024physics}. Instead, there are calls for studying the ``relative benefits of different active learning methods and the most effective means of implementation''~\cite[p. 8320]{wieman2014large}. Previous research has separately found a wide variation in instructor modifications of active learning methods~\cite{dancy2010pedagogical,turpen2009not,wood2016characterizing} and that these different implementations can exhibit a range of student learning gains~\cite{hake1998interactive,andrews2011active}.


This study focuses on the most widely used active learning pedagogy: Peer Instruction~\cite{dancy2024physics}. Peer Instruction typically involves the instructor asking several conceptual questions (i.e., clicker questions) during class, where students first individually answer the question and then discuss the question in small groups of nearby peers before re-answering the question and hearing the instructor explain the correct solution~\cite{mazur1997peer}. Many studies have documented the benefits of Peer Instruction to student conceptual learning~\cite{fagen2002peer, crouch2001peer,lasry2008peer} and problem solving skills~\cite{giuliodori2006peer,cortright2005peer}. 

Researchers have also characterized what Peer Instruction looks like in practice using direct classroom observations. In two different studies, the authors found that even for introductory physics instructors who use Peer Instruction in similar courses at the same institution, instructional practices varied significantly~\cite{turpen2009not,wood2016characterizing}. For example, different instructors spent varying amounts of class time asking clicker questions and explaining the solutions to these questions.


These studies importantly document different implementations of Peer Instruction; however, they do not relate the implementations to student outcomes. In other words, we lack an understanding of what specific features of implementation of Peer Instruction are beneficial to student learning. The studies examining implementation of Peer Instruction also take place at a single institution~\cite{turpen2009not,wood2016characterizing}, capturing limited instructional contexts and student populations. In this study, we directly compare instructor implementations of Peer Instruction to student conceptual learning using a diverse data set from seven instructors at six institutions. We aim to address the following research question: how does the implementation of Peer Instruction relate to student conceptual learning?


\section{Methods}

\begin{table*}[tb]
  \caption{Summary of institution types, number of students enrolled (and number of students with matched concept inventory responses), and concept inventories for the participating instructors. R1 indicates ``Very High Research Spending and Doctorate Production'' and RCU indicates ``Research Colleges and Universities,'' as determined by the Carnegie Classifications of Research Activity from 2025. 
  \label{courseinfo}}
  \begin{ruledtabular}
    \begin{tabular}{llll}
 Instructor & Institution Type  & No. of (Matched) Students &  Concept Inventory \\ 
 \hline
 A &  R1, Private, PhD-granting, Northeastern U. S. & 201 (106) & Mechanics Baseline Test~\cite{hestenes1992mechanics} \\ 
 B &  R1, Private, PhD-granting, Northeastern U. S. & 576 (389) & Mechanics Baseline Test \\ 
 C &   Public, Master's-granting, Northeastern U. S. & 28 (21) & Force Concept Inventory (FCI)~\cite{hestenes1992force}\\ 
 D &  R1, Private, PhD-granting, Northeastern U. S. & 81 (60) & Half-Length FCI~\cite{han2016experimental} \\ 
 E & Private, Master's-granting, Northeastern U. S. & 20 (14) & Force Concept Inventory\\ 
 F & RCU, Public, Master's-granting, Western U. S. & 94 (52) & Force Concept Inventory\\ 
 G & Private, PhD-granting, Women's college, Southeastern U. S. & 17 (15) & Gender-FCI~\cite{mccullough2001gender} \\ 
    \end{tabular}
  \end{ruledtabular}
\end{table*}

\subsection{Data collection}

As part of a national research project, we recruited seven instructors of introductory mechanics courses within the United States who self-reported using Peer Instruction (Table~\ref{courseinfo}). Each instructor collected two data sources: (1) pre- and post-semester concept inventories and (2) video recordings of three class sessions. We asked instructors to choose an existing concept inventory to give to their students such that the assessment of conceptual learning aligned with the topics taught in their particular course. Concept inventory data from all seven instructors had more than 50\% of enrolled students with matched responses (i.e., completing both the pre- and post-semester concept inventory). For the video recordings, instructors used Zoom or a camera that we sent to them via mail to record three consecutive class sessions, capturing a ``typical week'' in the course~\cite{stains2018anatomy}. All seven courses were held in person. 

\begin{table}[b]
  \caption{Summary of FILL+ codes for each interactivity level. Full definitions can be found at Ref.~\cite{kinnear2021developing}. \label{fillcodes}}
  \begin{ruledtabular}
    \begin{tabular}{ll}
 Interactivity Level & Codes \\ 
 \hline
  Interactive & Feedback (FB)\\
  & Student Discussion (SD)\\
   & Student Thinking (ST)\\

   & Class Question (CQ) \\

 Vicarious interactive & Student Question (SQ) \\
 & Lecturer Question (LQ) \\
  & Student Response (SR) \\
 & Lecturer Response (LR)\\
   Non-interactive  & Lecturer Talk (LT) \\
   & Administrative (AD) \\

    \end{tabular}
  \end{ruledtabular}
\end{table}

\subsection{Data analysis}

We first calculated Hedges' \textit{g}---a standardized mean difference between students' pre- and post-semester concept inventory scores---using the matched student responses in each course ~\cite{cohen1992quantitative,gurevitch1999statistical}. Hedges' \textit{g} is an effect size with thresholds similar to Cohen's \textit{d}: values below 0.2 are considered small effects, values between 0.2 and 0.5 are considered medium effects, and values above 0.8 are considered large effects~\cite{cohen2013statistical}. Researchers have recommended using Hedges' \textit{g} over other measures of learning, such as normalized gain, because Hedges' \textit{g} can be compared across courses of different sizes and is less susceptible to biases favoring students with high pre-scores~\cite{BurkholderConceptInventory,NissenCohenD,turner2006calculating}. We compared the values and confidence intervals of Hedges' \textit{g} across courses to examine the extent to which they varied. We made these comparisons qualitatively (i.e., comparing whether or not the confidence intervals overlap with one another), as we did not aim to make statistical claims about differences in effect sizes between courses given the small sample size.

We then applied the Framework for Interactive Learning in Lectures (FILL+) to the 21 video recordings (three per instructor)~\cite{wood2016characterizing,kinnear2021developing}. FILL+ is an observation protocol that characterizes classroom activities in interactive lecture environments by documenting what the students and instructor(s) are doing at a given time. FILL+ is more suitable than other existing frameworks (e.g., the Classroom Observation Protocol for Undergraduate STEM~\cite{smith2013classroom}) for addressing our research question because it captures specific activities that typically occur in Peer Instruction classrooms~\cite{kinnear2021developing}. For example, FILL+ disentangles whether the instructor is posing a question to the whole class or asking an individual student to respond. FILL+ also captures the exact duration and chronological order of classroom activities: it is a continuous and mutually exclusive protocol. Finally, FILL+ classifies classroom activities into three categories indicating the extent to which the activity engages students in active learning (Table~\ref{fillcodes}):
\begin{enumerate}
    \item \textit{Interactive}: Activities involving interactions between students, between students and the instructor, or when a student interacts directly with the material (such as thinking about a clicker question individually). 
    \item \textit{Vicarious interactive}: Activities where most of the students are following along with the discussion, but neither actively participating nor passively listening.
    \item \textit{Non-interactive}: Activities with no interaction; students are passively listening.
\end{enumerate}

The first and second authors individually worked through the FILL+ training~\cite{Smith_Anderson_Gant_Kinnear_2024} and then independently applied the framework to one of the video recordings from this study. The intraclass correlation coefficient, a measure of interrater reliability, was 0.88 after the first round of coding, indicating sufficient agreement by the two authors~\cite{koo2016guideline}. The first author proceeded to code all of the remaining video recordings. 

We characterized the various implementations of Peer Instruction in two ways, both of which aggregated the three video recordings for each instructor. First, similar to prior work~\cite{wood2016characterizing}, we determined the fraction of total observed class time spent on each FILL+ code to determine the extent to which each instructor implemented different levels of interactivity. Second, we calculated the average number of codes per minute to measure the frequency with which instructors switch between different classroom activities (e.g., if an observation included the sequence of codes LT, CQ, ST, FB, CQ, and SD, this would be counted as six codes). 
We qualitatively compared these two measures to student conceptual learning (i.e., effect sizes) for each instructor to identify possible relationships between implementation and student learning. Due to our small sample size, we did not conduct any statistical tests.


\section{Results}

\subsection{Variation in student conceptual learning in different implementations of Peer Instruction}

\begin{figure}[t]
    \includegraphics[width=0.48\textwidth]{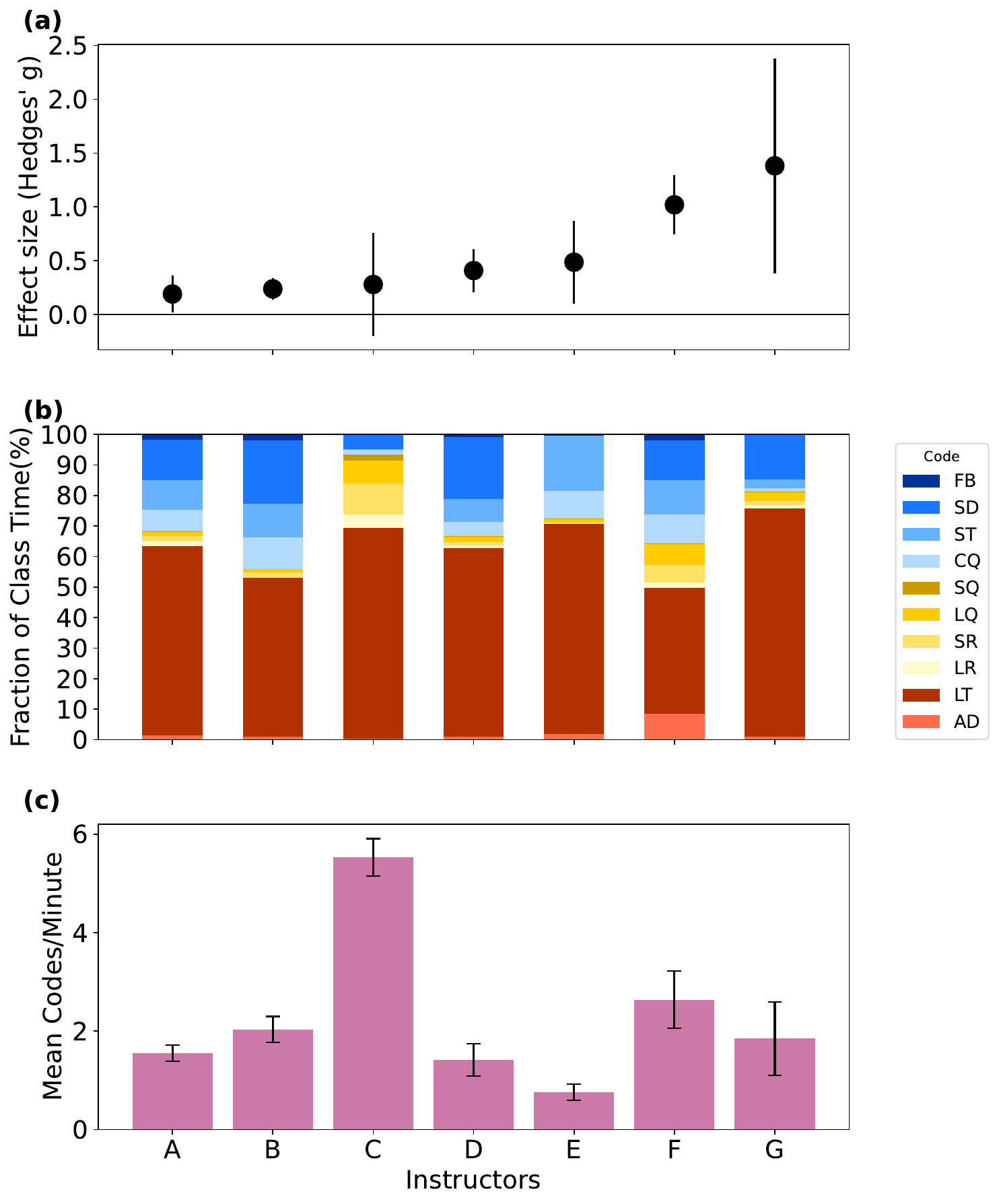}
    \caption{(a) Effect sizes for concept inventory scores. Points represent Hedges' \textit{g} values and error bars indicate 95\% confidence intervals. Positive values indicate increases in student scores from pre- to post-semester. (b) Bar charts indicating the fraction of observed class time spent on each FILL+ code (Table~\ref{fillcodes}). Blue colors indicate interactive codes, yellow colors indicate vicarious interactive codes, and red colors indicate non-interactive codes. (c) Mean number of FILL+ codes per minute of observed class time. Values (error bars) indicate means (standard deviations) across each instructor's three video recordings.}\label{fig1}
\end{figure}

All but one implementation of Peer Instruction exhibited positive gains in student conceptual learning (the confidence intervals of Hedges' \textit{g} values do not overlap with zero for all instructors except for Instructor C in Fig.~\ref{fig1}a). At the same time, there is a wide range in the magnitudes of these gains across implementations: the effect sizes range from 0.2 (Instructor A) to 1.4 (Instructor G).  Many of the confidence intervals of the effect sizes overlap with one another, indicating indistinguishable effects across implementations. There are, however, a few exceptions to this pattern: 
\begin{itemize}
    \item Instructor F exhibits larger conceptual learning gains than Instructors A, B, and D, and 
    \item Instructor G exhibits larger conceptual learning gains than Instructors A and B.
\end{itemize}

\subsection{Relationship between implementation of Peer Instruction and student conceptual learning}


In some regards, there is not much variation in the extent to which each instructor spends class time on different levels of interactivity as measured by FILL+. Six out of the seven instructors (all but Instructor F), for example, spend more than 50\% of class time on non-interactive activities, such as the instructor lecturing to the students (this is not unexpected in a Peer Instruction classroom, as even high fidelity implementations have 50\% of class time or more spent on instructor lecturing~\cite{crouch2001peer,commeford_characterizing_2021}; shades of red in Fig.~\ref{fig1}b). There also does not appear to be a clear relationship between the amount of class time spent on non-interactive activities and student conceptual learning. Instructors F and G, who exhibit the largest conceptual learning gains, spend the least and most class time, respectively, on non-interactive activities. Likewise, there is variation in the percentage of class time spent on lecturing across instructors A through E (ranging from 53\% to 71\%), and these courses exhibited comparable learning gains.

With the exception of Instructors F and G, each instructor predominantly uses just one of vicarious interactive or interactive activities when they are not lecturing (majority of either yellow or blue codes when ignoring the red codes in Fig.~\ref{fig1}b). Instructors A, B, D, and E predominantly use interactive strategies, such as clicker questions involving individual student thinking and/or student discussions (shades of blue in Fig.~\ref{fig1}b). Instructor C predominantly uses vicarious interactive strategies, such as brief lecturer questions to the whole class (shades of yellow in Fig.~\ref{fig1}b). Instructors F and G, who exhibit the largest conceptual learning gains, use a more balanced combination of both interactive and vicarious interactive strategies (similar amounts of blue and yellow shades in Fig.~\ref{fig1}b). Numerically, the ratios of vicarious interactive to interactive activities (considering all codes in each interactivity level and ignoring non-interactive codes) for Instructors A, B, D, and E range from 0.07 to 0.1, the ratio for Instructor C is 3.6, and the ratios for Instructors F and G are 0.4 and 0.3, respectively.


We also measured the rate at which each instructor transitions between FILL+ codes (Fig.~\ref{fig1}c). In most cases, these code transitions are switches between student-instructor interactions (e.g., the instructor presenting a clicker question) and student-student interactions (e.g., discussing a clicker question with peers). Instructor C, who is the only instructor with conceptual learning gains indistinguishable from zero, transitions between different activities the most frequently. This may be due to the predominant use of vicarious activities in this class. Instructor C asked many short questions to the whole class, prompting the students to give a simultaneous, shout-out response. Instructors A, B, D, F, and G transition between codes a moderate amount and Instructor E transitions between codes the least frequently.


\section{Discussion and Conclusion}

We build on previous studies that examine variation in instructor implementation of Peer Instruction at a single institution~\cite{turpen2009not,wood2016characterizing} by collecting similar data from a diverse set of institutions and also directly relating specific features of implementation to student conceptual learning. For the seven instructors included in our study, we find that some, but not all, features of implementation may relate to student learning.


\subsection{Class time spent on non-interactive activities does not seem related to student learning} 

Lecturing was prominent in most of the Peer Instruction classrooms in our sample; however, there was still considerable amounts of interactivity. Indeed, the instructor with the least amount of interactivity (i.e., the most amount of non-interactivity) in our sample (Instructor G) spent more time on interactive engagement than instructors in traditionally-taught classrooms, who typically spend 10\% or less of class time on interactivity~\cite{georgiou2014does}. Surprisingly, however, we found no clear relationship between the fraction of class time spent on non-interactive activities and student conceptual learning gains for the seven instructors included in this study. This preliminary finding does not fully align with results of prior studies, which have found a positive correlation between the amount of class time spent on interactive engagement and student learning~\cite{prather2009national}. This difference may be due to the way classroom activity is measured. We used direct observation, while previous studies rely on instructors' self-reports of their instructional practices which may not be accurate~\cite{ebert2011we}. Future work that relates direct observation to student learning and includes more instructors than our study is necessary to substantiate our initial findings.



\subsection{Relative amounts of interactive and vicarious interactive activities may relate to student learning}

The relative fractions of class time spent on interactive versus vicarious interactive activities (when ignoring the non-interactive activities) may relate to student learning. We found that the two instructors with the largest learning gains used a more balanced combination of interactive and vicarious interactive activities than the other five instructors, who predominantly used only one strategy (either interactive or vicarious interactive). We note that vicarious activities span a shorter time duration by nature (e.g., instructor questions that ask for a quick, shout-out response from the whole class), so a one-to-one ratio of time spent on interactive versus vicarious interactive activities is unlikely, even if there are similar numbers of vicarious interactive and interactive activities.



Although a larger data set is needed to validate these preliminary patterns, one possible explanation for this relationship is that combining these different types of activities may allow the instructor to receive more detailed feedback from the students than shout-out responses or clicker questions alone. When instructors pose a question to the class and one or a few students describe their thinking, the instructor may better understand different students' reasoning and clarify any misconceptions in real time~\cite{wood2016characterizing}. Future research should further examine the benefits, and reasons for these benefits, of vicarious interactive versus interactive activities. 




\subsection{Frequency of activities may relate to student learning}

We observed that the only instructor with student learning gains indistinguishable from zero had a much higher rate of transitioning between different activities than the other six instructors. As mentioned earlier, this pattern could be tied to the vicarious nature of the activities in this class. It may also be that students in this class were not given enough time to process what was happening at particular moments during class and/or had trouble following along with what they were supposed to be doing (i.e., listening or answering a question). We recommend for future research to further relate transitions between activities or interactivity levels to student learning by disentangling the frequencies of certain types of transitions (i.e., between specific pairs of FILL+ codes)~\cite{hora2014remeasuring}.

\subsection{Other aspects of implementation}

We acknowledge that classroom features beyond what we observed using the FILL+, such as class size and classroom layout, may also relate to student learning. Indeed, prior research suggests that there are challenges to incorporating active learning strategies in large classrooms and that students in small-enrollment active learning classes tend to have larger conceptual learning gains than students in large-enrollment active learning classes~\cite{freeman2014active,murdoch2002active,drinkwater2014managing}. In this study, however, class size and classroom layout do not seem related to student learning. Instructor F's class took place in a large lecture hall with 94 students, and exhibited larger learning gains than classes with smaller enrollment that took place in classrooms with small tables set up for group work. This finding corroborates a previous study comparing two identically-taught introductory biology courses, one held in a classroom set up for group work and one held in a traditional lecture hall, which found that student learning did not vary across layouts~\cite{stoltzfus2016does}. Peer Instruction, therefore, has the potential to be beneficial in a range of class sizes and classroom layouts. Other factors, such as the types of activities measured in this study, may better explain variation in student conceptual learning gains across Peer Instruction implementations. 


\subsection{Limitations and future work}

Our study illuminates variation in implementation of Peer Instruction across instructors at different institutions and directly relates implementation to student learning as measured from concept inventory scores. We only observed in-class activities during lectures, however, and other aspects of these courses, such as the discussion and laboratory sessions and homework assignments, may impact student understanding. FILL+ also does not capture the full variety of classroom activities that may be related to student learning. The framework, for example, does not tease apart demonstrations, simulations, and other student activities such as performing a physics-related dance (as we observed in one of the classes in this study), though all of these were integrated into the general structure of Peer Instruction rather than a separate activity in our data (e.g., a clicker question asking students to make predictions about a demonstration). We recommend for future research to characterize non-lecture course components and/or further develop FILL+ to include other types of activities to address these limitations and identify the extent to which student learning is attributable to the nature of Peer Instruction itself or specific types of activities within it. Future work should also measure student outcomes beyond conceptual learning, such as attitudes and beliefs, to achieve a more holistic understanding of the impacts of different Peer Instruction implementations on students. Finally, similar analyses that include a larger and more diverse sample, such as courses outside of the United States and at non-Master's granting and non-PhD granting institutions, will help to build on this study by enabling statistical claims.

\acknowledgments{We thank the instructors and students who participated in our study. This work is supported by the National Science Foundation Grant Nos. 2111128 and 2139899, and the Cotswold Foundation Postdoctoral Fellowship at Drexel University.
}

\bibliography{bibfile} 

\end{document}